# Modal Analysis for a Waveguide of Nanorods Using the Field Computation for a Chain of Finite Length

**Yaghoob Rahbarihagh, Farid Kalhor, Jalil Rashed-Mohassel, and Mahmoud Shahabadi**

Photonics Research Lab, Center of Excellence on Applied Electromagnetic Systems, School of ECE, College of Engineering, University of Tehran, North Kargar Ave, Tehran, Iran.
yrahbari@ut.ac.ir, f.kalhor@ut.ac.ir, jrashed@ut.ac.ir, shahabad@ut.ac.ir

*Abstract* — The propagation of light along an infinite 2D chain of silver nanorods is analyzed and the dispersion for this waveguide is computed using field computation for a finite chain of nanorods. In this work, Generalized Multipole Technique is used for the analysis. This method calculates the imaginary and real parts of the propagation constant by exciting the chain in one end and observing propagation of modes along the chain far enough from the excitation. It is shown that a short chain of finite length is sufficient for the calculation of the phase constant while the attenuation constant requires a longer chain. Field distribution is depicted for even and odd modes and it is shown that in the simulated frequency range only two modes can be excited and can propagate along the waveguide.

## I. INTRODUCTION

The ability of confining electromagnetic fields below the diffraction limit has made plasmonic waveguides a promising candidate for integrated optics. After Takahara et al. [1] demonstrated the possibility of guiding electromagnetic energy below the diffraction limit, various structures have been proposed as plasmonic waveguides. Quinten et al. [2] were the first to introduce a chain of metallic nanoparticles as a waveguide. Since then, this waveguide has been studied in many researches [3-15].

Several computational methods have been used for the analysis of nanoparticle-chain waveguides. Dipole Approximation (DA), Finite-Element Method (FEM) and Finite-Difference Time-Domain (FDTD) are commonly used methods. Dipole approximation is easy to implement and accurate for structures in which spacing of nanoparticles is significantly larger than particle dimensions. This method cannot be used for structures in which $L/r < 3$, where L is the separation between particles and r is the radius of the particle [16, 17]. Moreover, for particles of an arbitrary shape, calculation of polarizability demands additional numerical efforts. Improvements to DA like considering retardation effect [18], quadrupole and higher-order multipoles effect [16, 19] and adding the effect of layered background [20-22] have been proposed, yet it is not commonly used for a waveguide comprising arbitrary shaped nanoparticles with small inter-particle distance. FDTD and FEM are also common tools for analyzing plasmonic waveguides [4, 23]. However, in plasmonic structures, at frequencies near the plasma resonant frequency, electromagnetic (EM) fields are mainly confined around particles. Therefore, for domain discretization methods, such as FDTD and FEM, a very fine mesh is needed to achieve acceptable accuracy. A comparison between domain discretization and boundary discretization methods can be found in [24]. According to [24], boundary discretization methods show higher precision and are less time consuming for 2D plasmonic structures.

Generalized Multipole Technique (GMT) is a boundary discretization method which has already been used for the modal analysis of nanoparticle-chain waveguides [8, 11]. This method is applicable to structures with $L/r < 3$. Also, GMT is capable of analysis of a waveguide comprising arbitrary shaped nanoparticles. Nevertheless, GMT works with smaller matrices, which leads to less physical memory consumption compared with the domain discretization methods. Dispersion diagram for a 2D and 3D waveguide of nanoparticles is calculated in [8, 11] using GMT. In these researches, the propagation constant is calculated by finding the extrema of a cost function, like error or field intensity. Finding extrema of this function requires calculation of the cost function at different frequencies, which can be time consuming. However, the extrema of the function can be affected by changes in the field distribution of the modes or coupling among modes for different frequencies. Moreover, finding the attenuation constant needs extra calculation.

An improved modal analysis is presented in this work. The EM field distribution in a finite chain of nanorods is analyzed and the complex wavenumber using the complex value of the EM field at certain sampling points is calculated. The propagation of the EM field in a finite chain of nanorods is computed using GMT formulation.

## II. GENERALIZED MULTIPOLE TECHNIQUE FORMULATION FOR 2D NANOSTRUCTURES

Generalized Multipole Technique is a frequency-domain method for solving Maxwell's equations after subdividing the solution domain into homogeneous subdomains [25]. The EM field in each subdomain is expanded in terms of the EM field generated by a number of multipoles placed outside the subdomain. The unknown amplitude of the multipoles are calculated satisfying boundary conditions with minimum error.

For a two-dimensional z-invariant problem, the z-components of the electric and magnetic field at a given point $\vec{r}$ generated by $N_l$ clusters of multipoles can be expanded as:

$$E_z(\vec{r}) = \sum_{l=1}^{N_l} \sum_{n=-N}^{N} \frac{k^2 C_{ln}}{j\omega\epsilon} H_n^{(2)}\left(k|\vec{r}_l|\right) e^{jn\varphi_l}, \quad (1a)$$

$$H_z(\vec{r}) = \sum_{l=1}^{N_l} \sum_{n=-N}^{N} \frac{k^2 D_{ln}}{j\omega\mu} H_n^{(2)}\left(k|\vec{r}_l|\right) e^{jn\varphi_l}, \quad (1b)$$

in which the $l$-th cluster containing $N$ multipoles is located at $\vec{r}_l$. The coefficients $C_{ln}$ and $D_{ln}$ are the amplitudes of $TE_z$ and $TM_z$ multipoles, respectively, $k$ is the wave-number of the subdomain in question and $\varphi_l$ is the angle at which location $\vec{r}$ is seen by the multipoles placed at $\vec{r}_l$. There are a total of $2N_l(2N+1)$ multipoles.

In a finite chain composed of m nanorods, as shown in Fig. 1, there are m+1 subdomains. For each subdomain, the z-components of the EM field can be expanded using equations 1a and 1b with clusters placed outside of the subdomain. Fields of the subdomain $D_1$ are expanded by all the clusters represented by (+) and fields inside each nanorod are expanded by a set of clusters placed around it. Excitations can be placed at arbitrary positions and are represented by (∗).

Matching of the tangential field components on the boundaries is ensured by means of generalized point matching (GPM). Matching tangential magnetic and electric fields at matching points leads to the following system of equations:

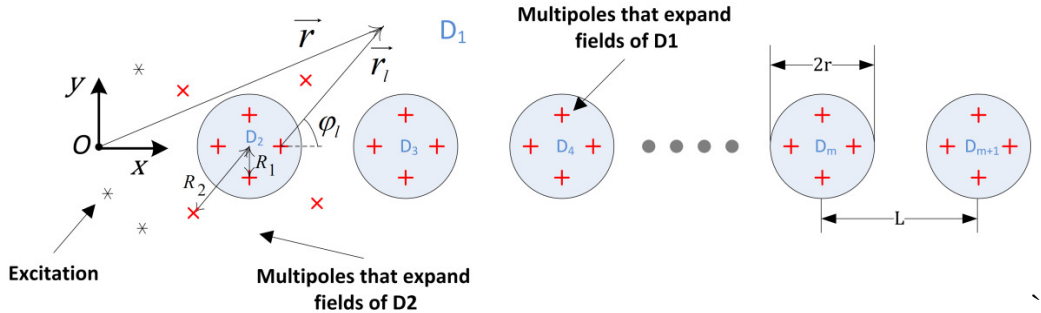

Fig. 1. Schematic of a finite chain containing *m* nanoparticles.

$$[A]_1 \begin{bmatrix} C \\ D \end{bmatrix}_1 + [B]_{exc} =$$

$$\begin{bmatrix} [A]_2 & 0 & \cdots & 0 \\ 0 & [A]_3 & \cdots & 0 \\ \vdots & \vdots & \ddots & \vdots \\ 0 & 0 & \cdots & [A]_{m+1} \end{bmatrix} \begin{bmatrix} \begin{bmatrix} C \\ D \end{bmatrix}_2 \\ \begin{bmatrix} C \\ D \end{bmatrix}_3 \\ \vdots \\ \begin{bmatrix} C \\ D \end{bmatrix}_{m+1} \end{bmatrix}, \quad (2)$$

where $\begin{bmatrix} C \\ D \end{bmatrix}_i$ are the unknown amplitudes of the multipoles expanding the EM field in the $i$-th subdomain, $[B]_{exc}$ is the tangential excitation fields at the matching points with the excitation placed in the first subdomain at (∗) locations in Fig. 1 and the matrices $[A]_i$ relate the unknown coefficients to the tangential fields at the matching points of the $i$-th subdomain.

Equation 2 presents an over-specified system. It can be reshaped to yield $[A]\begin{bmatrix} C \\ D \end{bmatrix} = [B]$ which may not have an exact solution. This equation is solved after computation of the pseudo-inverse of the $[A]$ matrix which minimizes the error defined by:

$$Error = \frac{\left\| [A]\begin{bmatrix} C \\ D \end{bmatrix} - [B] \right\|}{\|[B]\|}. \quad (3)$$

## III. MODAL ANALYSIS

In general, a metallic nanorod waveguide has various modes with different propagation constants. By its arbitrary excitation, a group of waveguide modes are excited and will propagate along the nanorods. If there is a dominant mode at each given frequency, other modes are attenuated according to a larger attenuation constant. Hence, by moving away from the excitation point, the amplitudes of the non-dominant modes decay faster than the amplitude of the dominant mode.

In a periodic structure for which only one mode propagates along the x-direction, the mode fields satisfy the Bloch condition:

$$\vec{f}(x+L, y) = \vec{f}(x, y) e^{-jkL}, \quad (4)$$

where $\vec{f}$ denotes the electric or magnetic field, $x$ represents the propagation direction, $L$ is the period of the structure, and $k = (\beta - j\alpha)$ is the propagation constant. In principle, by sampling the function $\vec{f}$ at various $x$, values one can determine the propagation constant, $k$.

For a finite but long waveguide and far from the excitation end, one may assume single-mode propagation of the dominant mode. Taking both the forward and backward propagating dominant mode into account, we may express $\vec{f}(x, y)$ as:

$$\vec{f}(x, y) = (Ae^{-jkx} + Be^{jkx})\vec{u}(x, y), \quad (5)$$

where $\vec{u}$ is a periodic function in $x$, $A$ and $B$ represent the amplitudes of the forward and backward propagating dominant mode, respectively. To determine $A$, $B$, and $k$, the total field must be sampled at least at three different locations. As a result of which, a system of three complex equations is obtained. Note that the three sampling points have a spacing of $L$ in the $x$-direction and have identical $y$-values, so the function $\vec{u}(x, y)$ has no influence on determination of the unknowns $A$, $B$, and $k$. Note that depending on the polarization of the excitation field, $TE_z$ or $TM_z$ modes are excited. Therefore, the function $\vec{f}$ is given either by equation 1a or 1b.

For a symmetric waveguide with respect to the propagation direction, i.e., the $x$-direction, modes can be classified as even or odd in terms of the electric or magnetic field component. If the waveguide is excited with even (odd) excitation, only even (odd) modes will be present along the finite chain. Thus, using an appropriate excitation, even (odd) dominant mode can be characterized.

Furthermore, taking symmetry along the $x$ and $y$-axis into consideration, we decrease the number of unknown coefficients and thus the boundary points.

Fig. 2 shows a finite chain excited symmetrically. MN represents the symmetry plane of the structure. Because of this symmetry, it will be adequate to solve Maxwell's equations only in one half of the structure. Other advantage of a symmetric excitation is that the amplitudes of forward and backward modes have to be equal in the middle of the structure. This eliminates one equation and one unknown coefficient.

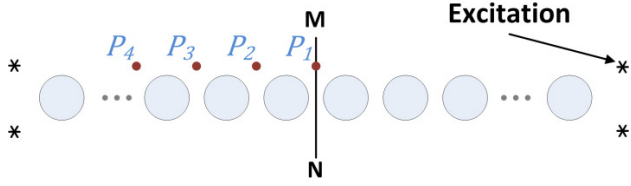

Fig. 2. A finite chain excited symmetrically.

There are two ways to verify the single-mode approximation discussed above. First, one can compare the field of the calculated mode and the total field at other points. Second, one can consider two existing modes in the waveguide. If the amplitude of the second mode is negligible compared to the first one, then the single-mode approximation leads to appropriate results. The following system of equations must be solved if two modes are propagating along the waveguide:

$$f_{P_1} = 2A + 2B,$$
$$f_{P_2} = A(e^{-jk_1L} + e^{jk_1L}) + B(e^{-jk_2L} + e^{jk_2L}),$$
$$f_{P_3} = A(e^{(-j2k_1L)} + e^{(j2k_1L)}) + B(e^{(-j2k_2L)} + e^{(j2k_2L)}),$$
$$f_{P_4} = A(e^{-j3k_1L} + e^{j3k_1L}) + B(e^{-j3k_2L} + e^{j3k_2L}),$$
(6)

where $A$ and $B$ are the amplitude of the first and second modes in the middle of the chain, respectively, whereas $k_1$ and $k_2$ are the propagation constants of these modes. The points $P_1$, $P_2$, $P_3$, and $P_4$ are shown in Fig. 2.

## IV. CONVERGENCE

Before we present the numerical results, we must investigate the convergence of our method. Convergence of wavenumber can be checked with respect to the defined error (equation 3) and the number of nanorods of the chain. Existence of a propagating mode, in a given frequency, is an essential prerequisite for convergence of the wavenumber. The next section shows (Fig. 6) in the simulated waveguide modes propagated in the $0.1 < L/\lambda < 0.13$ frequency range. Hence we investigate the convergence in this frequency range.

Fig.3 shows convergence of the computed wavenumber with respect to error. As shown in Fig. 3a, increasing the number of unknown coefficients decreases error. As error decreases, the propagation constant converges to its actual value (Fig. 3b). Increasing the number of unknown coefficients increases computational cost and time exponentially. Hence, a compromise should be made between accuracy and computational time.

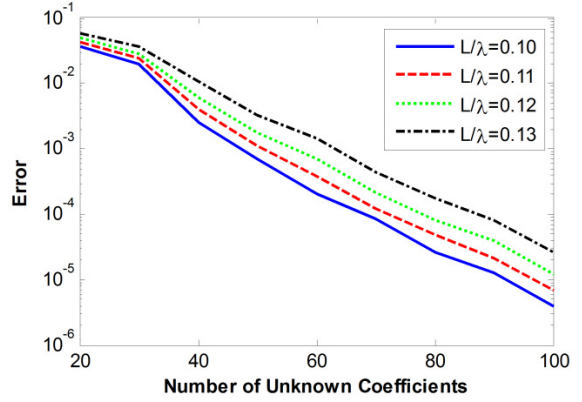

(a)

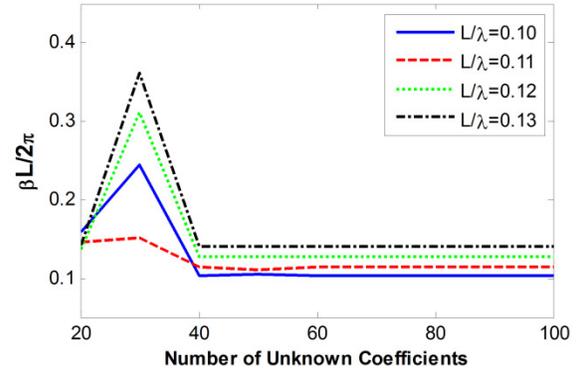

(b)

Fig. 3. Convergence of the wavenumber with respect to the number of unknown coefficients.

The second factor affecting convergence of the wavenumber is the number of nanorods in the chain. Increasing the chain length reduces the unwanted effects of the source and the non-dominant modes. Fig. 4 shows effects of the chain length on $\beta$ and $\alpha$. It shows that $\beta$ converges for a shorter chain whereas this is not the case for $\alpha$. It should be noted that the convergence at higher frequencies is faster.

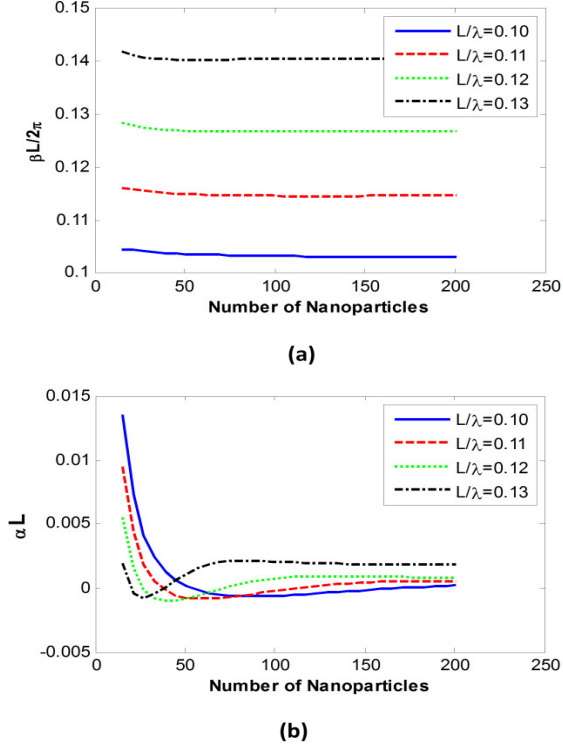

Fig. 4. Convergence of the wavenumber with respect to the number of nanorods. The geometrical and electrical parameters used for obtaining these diagrams are given in section 5.

## V. RESULTS

Using the method explained above, propagation of the EM field is calculated for a waveguide of nanorods. The waveguide comprises 100 silver nanorods with $r = 25nm$ and $d = 55nm$ (Fig. 1). Experimental data of [26] are used for the electrical permittivity of silver. The EM fields of both inside and outside of nanorods are expanded using six clusters of multipoles ($N_l = 6$) with $N = 2$ for even modes and nine clusters of multipoles ($N_l = 9$) with $N = 6$ for odd modes. The clusters of multipoles which expand the EM field outside and inside of the rods are placed at $R_1 = r/4$ and $R_2 = 2r$, respectively. This set of order and location of the multipoles leads to an error of less than 0.7% in the entire frequency range.

A typical EM field calculation showing propagation of the EM field along the chain is illustrated in Fig. 5. The inset of the Fig. 5 shows the amplitude of the magnetic field generated by the sources in free space. Note that an array with a null in the *x*-direction is used for the excitation of TE modes. Fig. 5 shows the propagation of the EM field along the chain for the same excitation. This figure clearly shows that the EM field is guided along the chain.

In order to calculate the dispersion diagram, the waveguide is stimulated by two sets of excitations; even excitation which only generates longitudinal modes and odd excitation for generating transverse modes. Fig. 6 shows the normalized amplitudes, $|A_e|, |A_o|$, of the dominant mode for even (longitudinal) and odd (transverse) modes respectively. As is shown in Fig. 6a, the even mode propagates along the chain in normalized frequencies below 0.145. In this figure, solid and dash lines show the amplitudes of the first ($|A|$) and the second ($|B|$) modes of equation 6, respectively. As the amplitude of the second mode is negligible in comparison with the first one, the single-mode approximation is acceptable in the normalized frequency range $0.06 < L/\lambda < 0.145$; thus, the non-propagating or the higher-order modes do not highly affect the results.

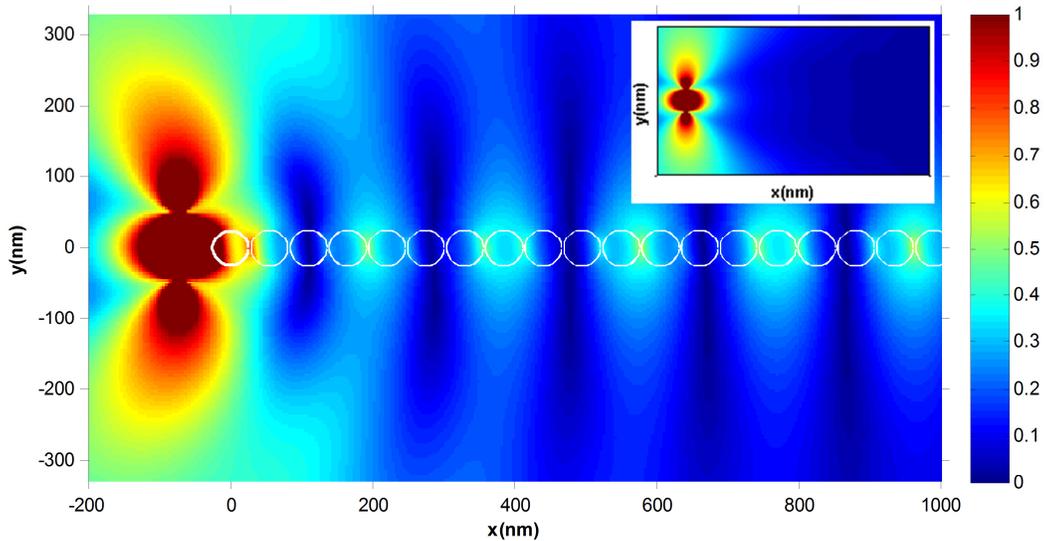

Fig. 5. Propagation of the EM field along the chain. This figure shows the normalized amplitude of the magnetic field for an even excitation. The inset shows the magnetic field of the sources in free space.

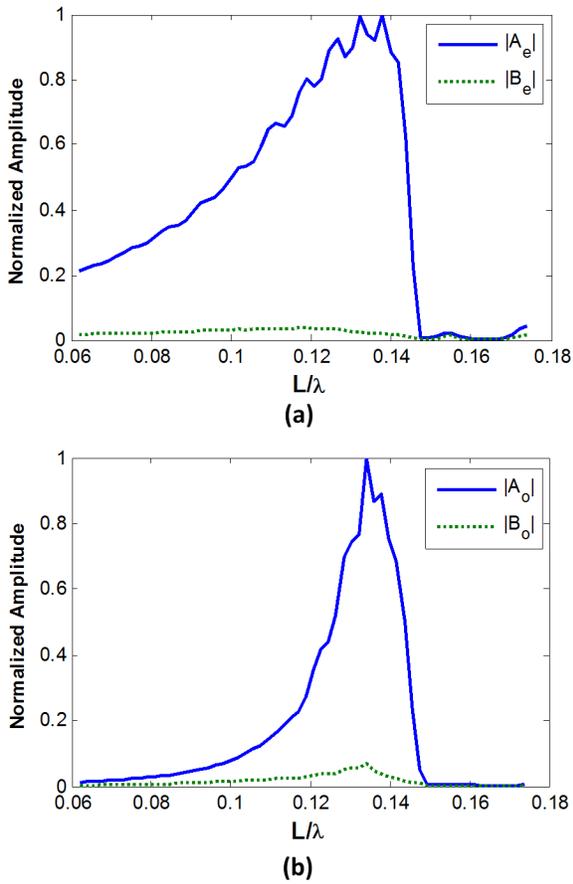

Fig. 6. Amplitudes of the first and second modes. (a) Even modes, (b) Odd modes.

Such as the even mode, the odd mode shows similar behavior in this frequency range. The higher normalized cutoff frequency is about 0.145. At this frequency the attenuation increases considerably. The lower normalized cutoff frequency for the odd mode is about 0.12, as will be discussed further in what follows.

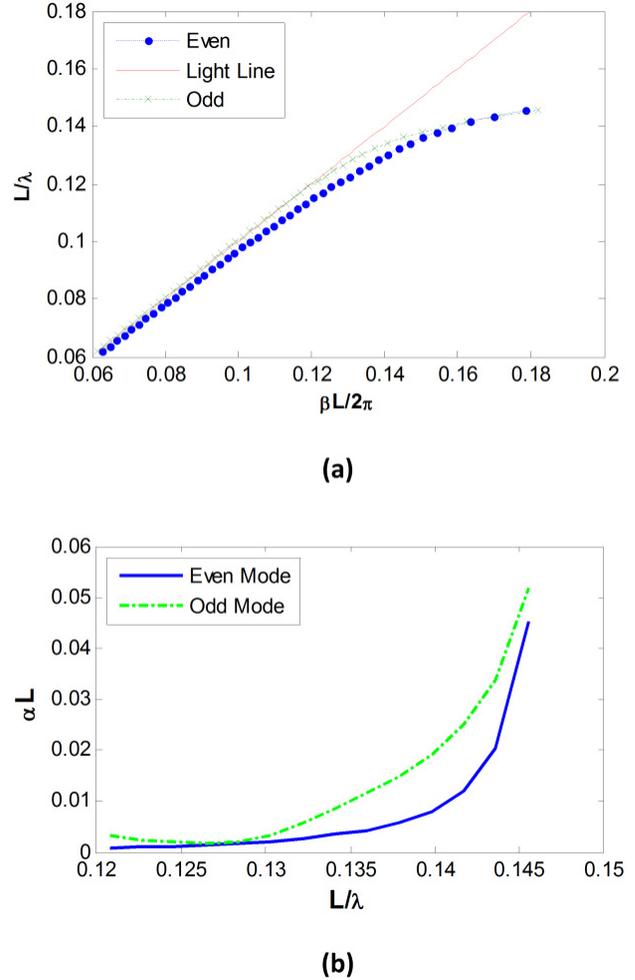

Fig. 7. Propagation (a) and attenuation (b) constants for the even and odd modes.

The propagation constant for the even and odd modes is calculated and depicted in Fig. 7a. The results are in agreement with the results reported by Talebi and Shahabadi [8] for the first even and odd modes. Also, the accuaracy of the GMT results is compared and validated with other techniques in [11, 24]. There is no higher-order modes or their extinction length is smaller than the length of the simulated chain. Note that $\alpha$ and $\beta$ are calculated simultaneously as a complex wave-number. Fig. 7b shows the attenuation constant $\alpha$ for even and odd modes. According to Fig. 4b, at lower frequencies, a longer chain is needed for the attenuation constant to converge. For a chain of

100 nanorods results for the normalized frequncies below 0.12 are not accurate, but follow the well-known behavior of the attenuation constant. The field distributions for these modes at a normalized frequency of $L/\lambda = 0.14$ are depicted in Fig. 8.

mode. Also, it shows that as the frequency increases, the EM field becomes more confined by the chain.

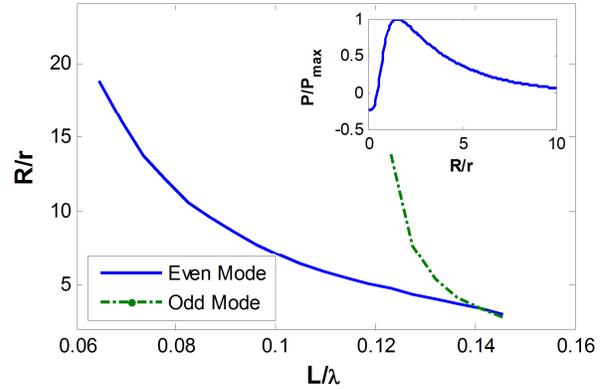

Fig. 9. Confinement of the EM fields for the even and odd modes.

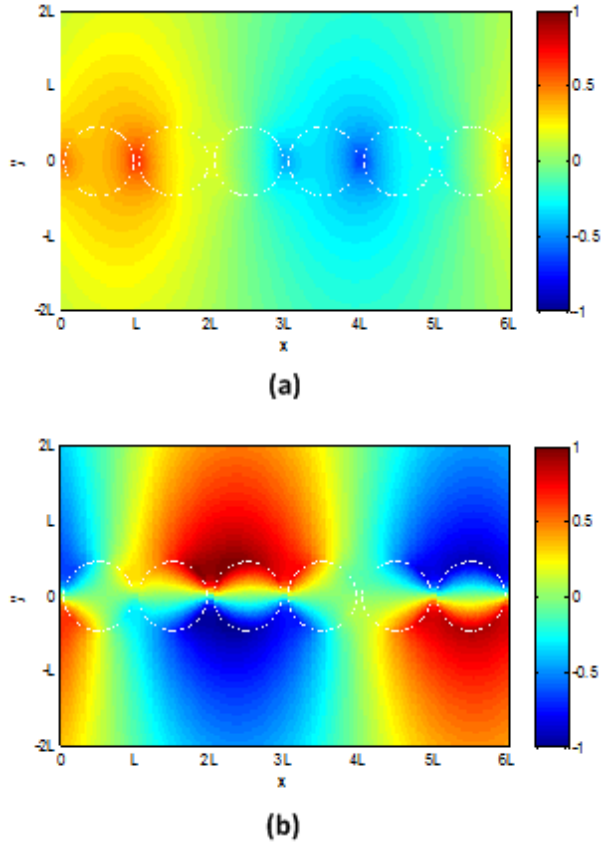

Fig. 8. Distribution of the normalized amplitude of the magnetic field at L/λ=0.14 (a) Even mode (b) Odd mode.

The confinement of the EM wave is an important characteristic of nanorod-chain waveguides. Let define $R/r$ as the normalized spacing from the waveguide axis at which the longitudinal component of the Poynting vector drops to half of its maximum value. Fig. 9 shows this normalized spacing $R/r$ as a function of the normalized frequency. The inset of this figure shows the normalized Poynting vector ($P/P_{max}$) in the propagation direction for different distances from the chain axis at the normalized frequency, $L/\lambda = 0.132$. This figure reveals that the even mode is more confined in comparison to the odd

## VI. CONCLUSION

In this paper, we calculated the dispersion diagram of a nanorod-chain waveguide using GMT and demonstrated that the propagation and attenuation constants can be obtained directly by observing propagation of modes along a finite chain. Using GMT, it is observed that a short chain of finite length is sufficient for the calculation of the phase constant. However the attenuation constant, $\alpha$, requires a longer chain. These parameters are calculated for two propagating modes in the normalized frequency range of $0.06 < L/\lambda < 0.18$. This method is more effective for calculation of the propagation and attenuation constants for modes with higher attenuation in comparison with previous report [8].

Although we have demonstrated this method for the modal analysis of a 2D waveguide of nanorods, it can be used for the analysis of 3D waveguide structures containing nanoparticles of an arbitrary shape.

## REFERENCES
[1] J. Takahara, S. Yamagishi, H. Taki, A. Morimoto, and T. Kobayashi, "Guiding of a one-dimensional optical beam with nanometer diameter," *Opt. Lett.* 22, 475–477 (1997).
[2] M. Quinten, A. Leitner, J. R. Krenn, and F. R. Aussenegg, "Electromagnetic energy transport via

linear chains of silver nanoparticles," *Opt. Lett*. 23, 1331–1333 (1998).

[3] M. L. Brongersma, J. W. Hartman, and H. A. Atwater, "Electromagnetic energy transfer and switching in nanoparticle chain arrays below the diffraction limit," *Physical Review B* 62, R16356–R16359 (2000).

[4] S. Maier, P. Kik, and H. Atwater, "Optical pulse propagation in metal nanoparticle chain waveguides," *Physical Review* B 67, (2003).

[5] R. A. Shore and A. D. Yaghjian, "Travelling electromagnetic waves on linear periodic arrays of lossless spheres," *Electronics Letters* 41, 578 – 580 (2005).

[6] A. Alù and N. Engheta, "Theory of linear chains of metamaterial/plasmonic particles as subdiffraction optical nanotransmission lines," *Departmental Papers* (ESE) (2006).

[7] M. Sukharev and T. Seideman, "Coherent control of light propagation via nanoparticle arrays," *Journal of Physics B: Atomic, Molecular and Optical Physics* 40, S283–S298 (2007).

[8] N. Talebi and M. Shahabadi, "Analysis of the Propagation of Light Along an Array of Nanorods Using the Generalized Multipole Techniques," *Journal of Computational and Theoretical Nanoscience* 5, 711–716 (2008).

[9] A. Hochman and Y. Leviatan, "Rigorous modal analysis of metallic nanowire chains," *Optics Express* 17, 13561 (2009).

[10] Y. Hadad and B. Z. Steinberg, "Green's function theory for infinite and semi-infinite particle chains," *Phys. Rev*. B 84, 125402 (2011).

[11] S. M. Raeis Zadeh Bajestani, M. Shahabadi, and N. Talebi, "Analysis of plasmon propagation along a chain of metal nanospheres using the generalized multipole technique," *J. Opt. Soc. Am. B 28*, 937–943 (2011).

[12] D. Szafranek and Y. Leviatan, "A Source-Model Technique for analysis of wave guiding along chains of metallic nanowires in layered media," *Optics Express 19*, 25397 (2011).

[13] I. B. Udagedara, I. D. Rukhlenko, and M. Premaratne, "Complex-ω approach versus complex-k approach in description of gain-assisted surface plasmon-polariton propagation along linear chains of metallic nanospheres," *Phys. Rev. B 83*, 115451 (2011).

[14] B. Rolly, N. Bonod and B. Stout, "Dispersion relations in metal nanoparticle chains: necessity of the multipole approach," *J. Opt. Soc. Am. B 29*, 1012-1019 (2012).

[15] K. Sendur, "Optical aspects of the interaction of focused beams with plasmonic nanoparticles," Applied Computational Electromagnetics Society (ACES) Journal, vol. 27, no. 2, pp. 181-188, February 2012.

[16] S. Y. Park and D. Stroud, "Surface-plasmon dispersion relations in chains of metallic nanoparticles: An exact quasistatic calculation," *Phys. Rev. B 69*, 125418 (2004).

[17] C. Tserkezis and N. Stefanou, "Calculation of waveguide modes in linear chains of metallic nanorods," *J. Opt. Soc. Am. B 29*, 827–832 (2012).

[18] W. H. Weber and G. W. Ford, "Propagation of optical excitations by dipolar interactions in metal nanoparticle chains," *Phys. Rev. B 70*, 125429 (2004).

[19] D. Han, Y. Lai, K. H. Fung, Z.-Q. Zhang, and C. T. Chan, "Negative group velocity from quadrupole resonance of plasmonic spheres," *Phys. Rev. B 79*, 195444 (2009).

[20] E. Simsek, "On the Surface Plasmon Resonance Modes of Metal Nanoparticle Chains and Arrays," *Plasmonics 4*, 223–230 (2009).

[21] D. Van Orden, Y. Fainman, and V. Lomakin, "Optical waves on nanoparticle chains coupled with surfaces," *Opt. Lett. 34*, 422–424 (2009).

[22] E. Simsek, "Full Analytical Model for Obtaining Surface Plasmon Resonance Modes of Metal Nanoparticle Structures Embedded in Layered Media," *Opt. Express 18*, 1722–1733 (2010).

[23] I. Tsukerman, "A new computational method for plasmon resonances of nanoparticles and for wave propagation," IEEE/ACES International Conference on Wireless Communications and Applied Computational Electromagnetics, pp. 909-912, Honolulu, Hi, April 2005.

[24] J. Smajic, C. Hafner, L. Raguin, K. Tavzarashvili, and M. Mishrikey, "Comparison of Numerical Methods for the Analysis of Plasmonic Structures," *Journal of Computational and Theoretical Nanoscience 6*, 763–774 (2009).

[25] C. Hafner, The Generalized Multipole Technique for ComputationalElectromagnetics (Artech House Publishers, 1990).

[26] P. B. Johnson and R. W. Christy, "Optical Constants of the Noble Metals," *Phys. Rev. B 6*, 4370–4379 (1972).

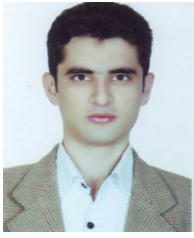
**Yaghoob Rahbarihagh** received the B.S. degree in electrical engineering from the University of Tehran, Iran, in 2009, and the M.S. degree in Communications Engineering (Fields and Waves) from the University of Tehran, Iran, in 2012. He was an engineering research associate at the Microwave and Photonic Research Labs at the University of Tehran. His research interests include photonics, plasmonics, and electromagnetics.

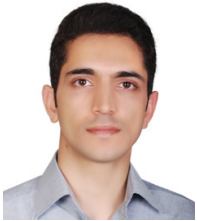
**Farid Kalhor** received the B.S. degree in electrical engineering from the University of Tehran, Iran, in 2009, and the M.S. degree in Communications Engineering (Fields and Waves) from the University of Tehran, Iran, in 2012. He was an engineering research associate at the Microwave and Photonic Research Labs at the University of Tehran. His research interests include photonics, plasmonics, and electromagnetics

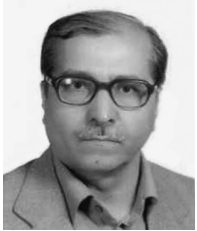
**Jalil Rashed-Mohassel** received the M.Sc. degree in electronics engineering from the University of Tehran, Tehran, Iran, in 1976 and the Ph.D. degree in electrical engineering from the University of Michigan, Ann Arbor, in 1982.

Previously, he was with the University of Sistan and Baluchestan, Zahedan, Iran. In 1994, he joined University of Tehran where he is teaching and performing research as a Professor in antennas, EM theory and applied mathematics. He served as the academic Vice-Dean, College of Engineering, General Director of Academic Affairs, University of Tehran and Chairman of the school of ECE, Principal member of Center of Excellence on Applied Electromagnetic Systems and the Director of the Microwave Laboratory.

Prof. Rashed-Mohassel was selected as the Brilliant National Researcher by the Iranian Association of Electrical and Electronics Engineers in 2007, and was the Distinguished Professor (2008–2009) in the 1st Education Festival, University of Tehran.

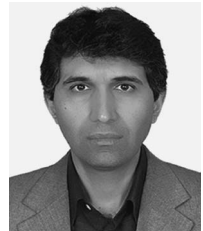
**Mahmoud Shahabadi** received the B.Sc. and MSc degrees from the University of Tehran, Tehran, Iran, and the Ph.D. degree from Technische Universitaet Hamburg-Harburg, Germany, in 1988, 1991, and 1998, respectively, all in electrical engineering.
Since 1998, he has been an Assistant Professor, Associate Professor and then a Professor with the School of Electrical and Computer Engineering, University of Tehran. From 2001 to 2004, he was a Visiting Professor with the Department of Electrical and Computer Engineering, University of Waterloo, Canada.

His research interests and activities encompass various areas of microwave and millimeter-wave engineering, as well as photonics. Computational electromagnetics for microwave engineering and photonics are his special interests.